\documentclass{aa}

\usepackage{graphicx,txfonts,natbib}

\begin{document}

\title{On using the CMB shift parameter in tests of models of dark energy}

\author{{\O}ystein Elgar{\o}y\inst{1}
        \and 
         Tuomas Multam\"{a}ki\inst{2}}

\offprints{{\O}. Elgar{\o}y}
\institute{Institute of theoretical astrophysics, University of Oslo, Box 1029, 0315 Oslo, Norway\\ 
\email{oelgaroy@astro.uio.no}
 \and 
 Department of Physics, University of Turku, FIN-20014 Turku, Finland \\
\email{tuomul@utu.fi}
}
\date{Received; accepted}

\abstract
{The so-called shift parameter is related to the position of the first acoustic 
peak in the power spectrum of the temperature anisotropies of the cosmic microwave background (CMB)
anisotropies. It is an often used quantity in simple tests of dark energy models. However, the 
shift parameter is not directly measurable from the cosmic microwave background, and its 
value is usually derived from the data assuming a spatially flat cosmology with dark matter 
and a cosmological constant.}
{To evaluate the effectiveness of the shift parameter as a constraint on dark energy models.}
{We discuss the potential pitfalls in using the shift parameter 
as a test of non-standard dark energy models.}
{By comparing to full CMB fits, we show
that combining the shift parameter with the position of the first acoustic peak in 
the CMB power spectrum improves the accuracy of the test considerably.}
{}
\maketitle

\section{Introduction}

Comparing cosmological models to current observational data can be cumbersome and 
computationally intensive. Large multi-dimensional parameter spaces cannot be
probed by grid-based methods but more sophisticated approaches are required, for example 
Monte Carlo Markov Chains (MCMC) \cite{gamerman,lewis}. Of the most commonly used 
cosmological data sets available today, cosmic microwave background (CMB) and 
large-scale structure (LSS) observations in particular require computational effort in parameter 
estimation. Given computing time and patience, this is not a problem, at least in principle, 
when testing models in which the evolution 
of linear density perturbations is well understood and hence calculable.
However, for several of the more imaginative dark energy models the situation is more 
complicated.  
In cases where the model is specified by an action, like the DGP model 
\cite{dvali}, 
one should in principle be able to set up the equations for linear perturbations, 
but in practice this has turned out to be difficult and it is only 
recently that progress in this direction has been made in this particular case
\cite{koyamab,koyamaa,sawicki,song}.
And even though the equations are known, they may be so complicated to 
treat numerically as to make it practically impossible to explore the 
parameter space of the model in MCMC. 
In addition, there are a large number of dark energy models, based on
phenomenological considerations, that lack the detail to allow one to 
proceed with well-defined calculations. Examples of such models include
the various proposed modifications of the Friedmann equation, where
the model is simply not specified well enough to allow the calculation 
of the density perturbations \citep[see e.g.][]{freese,gondolo}. 
The justification of such models may be questioned, but
the state of our understanding of dark energy argues for keeping an open mind. 
One would like to have some means of testing both groups of models, 
incorporating as much empirical information as possible, but avoiding 
the need to calculate the behaviour of density perturbations.  
In practice, this means restricting the observational tests to those 
involving the age and distance scale, in particular the luminosity 
distance-redshift relationship as probed by supernovae of type Ia (SNIa).  Important as 
the supernova data are, they are still not very restrictive if one allows 
for e.g. non-zero spatial curvature or a time-varying equation of state 
for dark energy \cite{riessa,perlmutter,tonry,barris,riessb,astier,clocchiatti,woodvasey,miknaitis,davis}.  

To tighten up constraints on dark energy models, 
a common approach is therefore to include additional information about 
the distance scale from the CMB in the form of the so-called 
shift parameter that is related to the position of the first acoustic 
peak in the power spectrum of the temperature anisotropies 
\cite{efstathiou}.  
Recently, after the baryon acoustic oscillations (BAO) where observed in the 
SDSS Luminous Red Galaxy sample \cite{eisenstein}, 
it has also become common to include the 
information about the angular scale of the oscillations \citep{davis,wright}.  
What one should bear in mind, however, is that these distance scales 
are not directly measured quantities, but are derived from the observations 
by assuming a specific model, usually the flat $\Lambda$CDM model or a 
slight variation thereof.  Care needs to be exercised when using these 
derived quantities to test more exotic dark energy models.

Here we consider the shift parameter in more detail by comparing its predictions to those 
obtained from full CMB fits for different types of cosmological models. We identify the 
limitations of using such a measure and advocate using a combination of the acoustic peak 
scale along with the shift parameter as a more accurate probe of the CMB power spectrum.
Such a combination is quick and easy to implement and should be 
included in tests of dark energy models where it is either cumbersome or unknown how to 
calculate the full CMB and matter power spectrum.  

\section{Theory}
The use of the shift parameter as a probe of dark energy 
is based on the observation that different models will have an 
almost identical CMB power spectra if all of the following criteria are satisfied \cite{efstathiou}:
$\omega_{\rm c}=\Omega_{\rm c}h^2$ and $\omega_{\rm b}=\Omega_{\rm b}h^2$ are equal, primordial fluctuation spectrum is unchanged,
and the shift parameter,
\begin{equation}
{\cal R}=\frac{\omega_{\rm m}^{1/2}}{\omega_{\rm k}^{1/2}}{\rm sinn}_k
(\omega_{\rm k}^{1/2}y),
\label{ebr}
\end{equation}
where $\textrm{sinn}(x)=\{\sin(x),x,\sinh(x)\}$ for $k=+1,0,-1$ respectively,
with 
\begin{equation}
y=\int_{a_{\rm r}}^1\frac{da}{\sqrt{\omega_{\rm m} a+\omega_{\rm k} a^2+
\omega_\Lambda a^4+\omega_{\rm Q} a^{1-3w}}}
\label{ebry}
\end{equation}
is constant. In this original definition of the shift parameter, the universe is 
considered to be filled with matter (dark and baryonic), $\omega_{\rm m}
=\Omega_{\rm m}h^2$, 
$\omega_{\rm b}$, curvature, $\omega_{\rm k}=\Omega_{\rm k}h^2$,
cosmological constant $\omega_\Lambda=\Omega_\Lambda h^2$ and a 
dark energy component 
$\omega_{\rm Q}=\Omega_{\rm Q}h^2$ with a
constant equation of state $w$. The density parameter $\Omega_{\rm i}$ 
is the ratio of the present-day density of component $\rm i$ to the 
density of a spatially flat universe, $\rho_{\rm c} = 3H_0^2/8\pi G$, 
and $h$ is the dimensionless Hubble constant defined by 
$H_0 = 100h\;{\rm km}\,{\rm s}^{-1}\,{\rm Mpc}^{-1}$. 
Integration is carried out from recombination, $a_{\rm r}$, until today $a=1$.

In a spatially flat universe ($k=0$), the shift parameter reduces to
\begin{equation}
{\cal R}=\sqrt{\Omega_{\rm m}}\int_0^{z_{\rm r}}\frac{dz}{E(z)},
\label{shift}
\end{equation}
where $E(z)\equiv H(z)/H_0$ and $H(z)$ is the Hubble parameter.
In this form the shift parameter has been used in a number of works  
\cite{lazkoz,wang,amarzguioui,elgaroy}.

The sound horizon at recombination for three
massless neutrinos is given by \cite{efstathiou}
\begin{eqnarray}
r_{\rm s}&=&\frac{c}{\sqrt{3}H_0}\Omega_{\rm m}^{-1/2}\int_0^{a_{\rm r}}
\frac{da}{\sqrt{(a+a_{\rm eq}))(1+R(a))}} \nonumber \\ 
&\approx& \frac{19.8\;{\rm Mpc}}{\sqrt{\omega_{\rm b}\omega_{\rm m}}}\ln\left(\frac{\sqrt{R(a_{\rm r})+R(a_{\rm eq})}
+\sqrt{1+R(a_{\rm r})}}{1+\sqrt{R(a_{\rm eq})}}\right),
\label{rs}
\end{eqnarray}
where $R(a)=30496\omega_{\rm b} a$ and $a_{\rm eq}=1/(24185\omega_{\rm m})$ 
and the recombination redshift can be calculated by using the fitting 
formulae \cite{hu}:
\begin{eqnarray}
z_{\rm r} & = & 1048(1+0.00124\omega_{\rm b}^{-0.738})(1+g_1\omega_{\rm m}^{g_2})\label{zrfit}\\
g_1 & = & 0.0783\omega_{\rm b}^{-0.238}/(1+39.5\omega_{\rm b}^{0.763})\nonumber\\
g_2 & = & 0.560/(1+21.1\omega_{\rm b}^{1.81}).\nonumber
\end{eqnarray}
The $m$th Doppler peak has the comoving wave number \cite{hu}
$m\pi=k_m r_s(a_{\rm r})$,
and hence the location of the first peak in multipole space is 
approximately given by
\begin{equation}
\ell _{\rm a}\approx \pi \frac{d_{\rm A}(z_{\rm r})}{r_{\rm s}(a_{\rm r})},
\label{peak1}
\end{equation}
where $d_A={\cal R}/\omega_{\rm m}^{1/2}$ is the angular diameter distance.
Rewriting Eq. (\ref{peak1}), we have
\begin{equation}
\ell _{\rm a}\approx 151 \omega_{\rm b}^{1/2}{\cal R}
\left(\ln\left(\frac{\sqrt{R(a_{\rm r})+R(a_{\rm eq})}
+\sqrt{1+R(a_{\rm r})}}{1+\sqrt{R(a_{\rm eq})}}\right)\right)^{-1}.
\label{peak}
\end{equation}
The location of the first peak is hence a combination of the
shift parameter and the size of the sound horizon at recombination, as
is expected on physical grounds. Even though the relation between 
$\ell _{\rm a}$ 
and ${\cal R}$ is linear, the two parameters are not degenerate and in fact
complement each other well in constraining models, as is shown later.




The best fit value calculated from the WMAP-team provided MCMC chains for the 
shift parameter in the standard flat $\Lambda$CDM model is
\begin{equation}
{\cal R} = 1.71^{+0.03}_{-0.03},
\label{wmap3shift}
\end{equation}
which is in good agreement with Wang \& Mukherjee (2006). This result is practically
equal for the $\Lambda$CDM and $w$CDM models with or without dark energy perturbations.
The acoustic peak position as measured by Eq. (\ref{peak}) calculated
from the same data is
\begin{equation}
\ell_{\rm a}=303.6^{+1.1}_{-1.2}.
\label{wmap3la}
\end{equation}

One should always bear in mind the conditions for the shift parameter 
to be applicable.  If one wants to use the shift parameter as a constraint 
on a dark energy model, then first of all the distribution of the 
shift parameter has to be derived from the CMB data.  This cannot be 
done without assuming a model.  Since one is only looking for a 
constraint on the expansion history of the universe, it is easy to forget 
that one is also making assumptions about the primordial power spectrum 
of density fluctuations, since these form the basis for calculating the 
CMB anisotropies.  In effect, one is therefore always making implicit 
assumptions about inflation, even though what one wants to test 
is the kinematics of the dark energy model under scrutiny.  

\begin{figure}[ht]
\begin{center}
\includegraphics[width=50mm,angle=0]{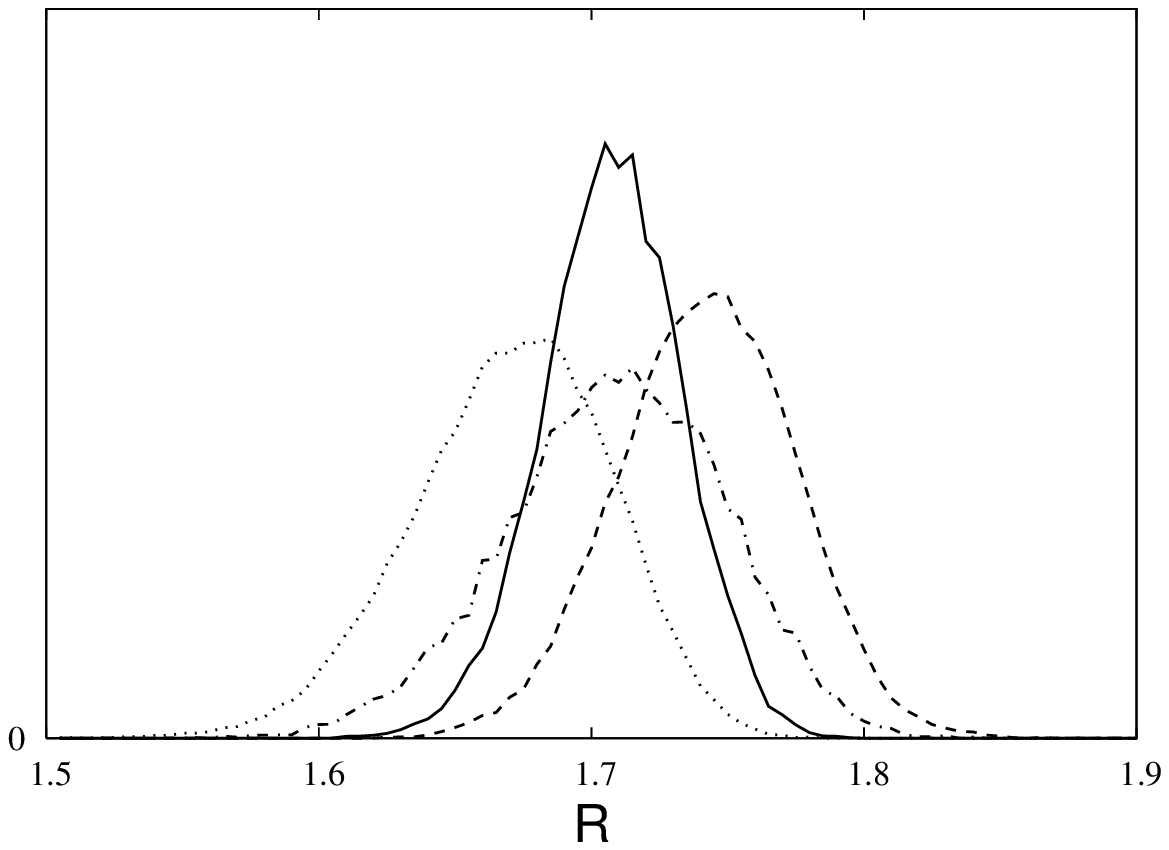}
\includegraphics[width=50mm,angle=0]{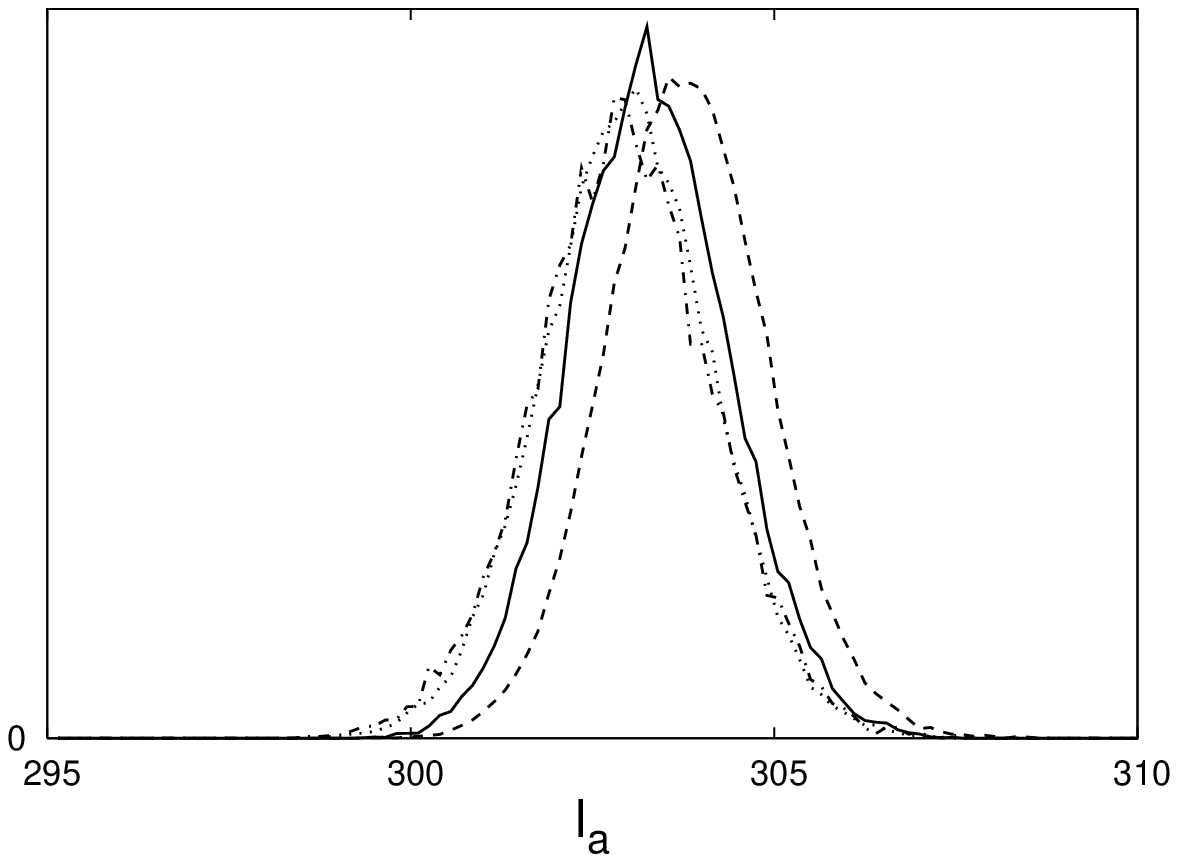}
\caption{The distribution of the shift parameter ${\cal R}$ and the 
acoustic scale $\ell _{\rm a}$ derived from MCMC chains with 
the WMAP data for four different types of primordial perturbations: power-law 
with no running scalar spectral index and no tensor modes (full line), power-law with tensor modes (dotted line), running spectral index and no tensor modes 
(dashed line), and both tensor modes and running spectral index (dot-dashed line).}\label{fig1}
\end{center}
\end{figure}

In order to demonstrate the significance of these underlying assumptions,
we show the distributions of the shift parameter in figure \ref{fig1} 
derived from MCMC chains for the $\Lambda$CDM model with four 
different primordial power spectra: the standard power-law version, 
power-law with running scalar spectral index, with tensor modes, and with 
both tensor modes and running scalar spectral index. The distributions 
for ${\cal R}$ are visibly different in the four cases.  
Therefore, whenever one uses the shift parameter one should be clear 
about the assumptions made in deriving its distribution from the 
CMB data.  The acoustic scale $\ell _{\rm a}$, also shown in figure \ref{fig1}, 
varies significantly less when changing the assumptions about the primordial 
power spectrum.  

\begin{figure}[ht]
\begin{center}
\includegraphics[width=50mm,angle=0]{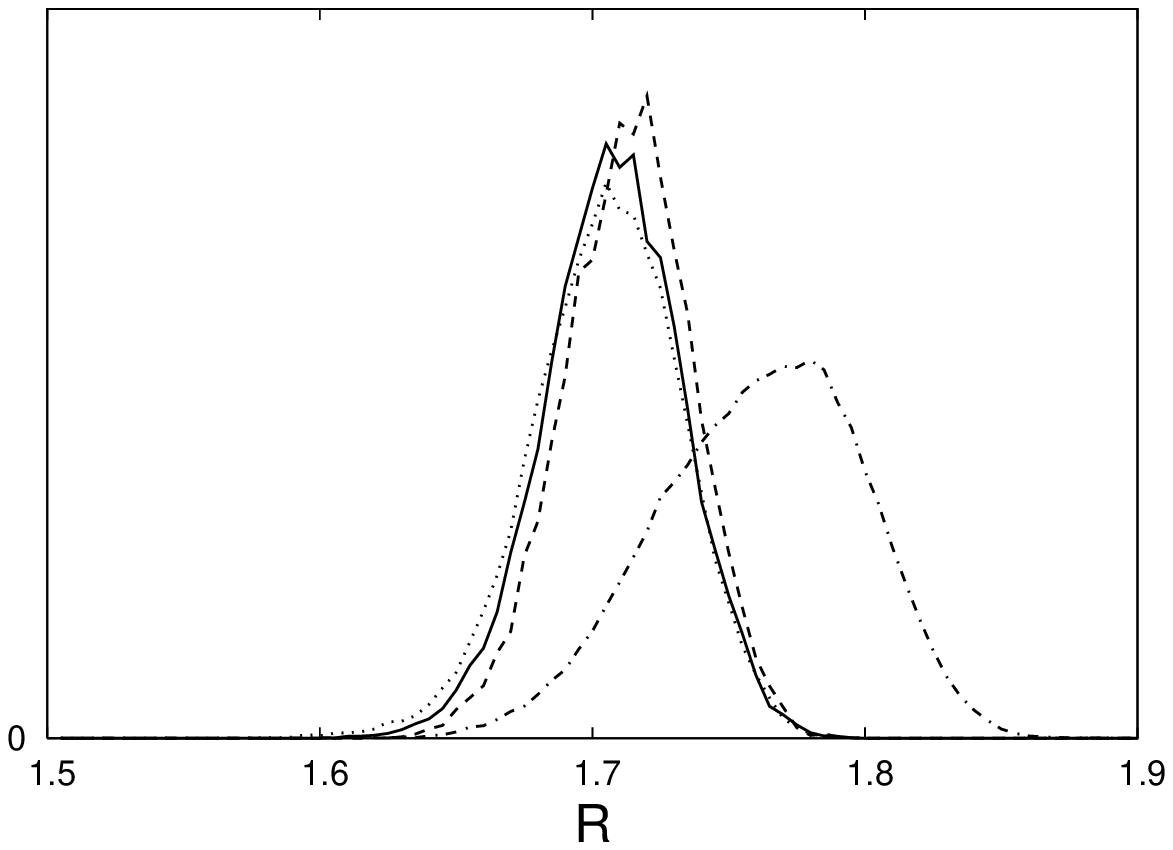}
\includegraphics[width=50mm,angle=0]{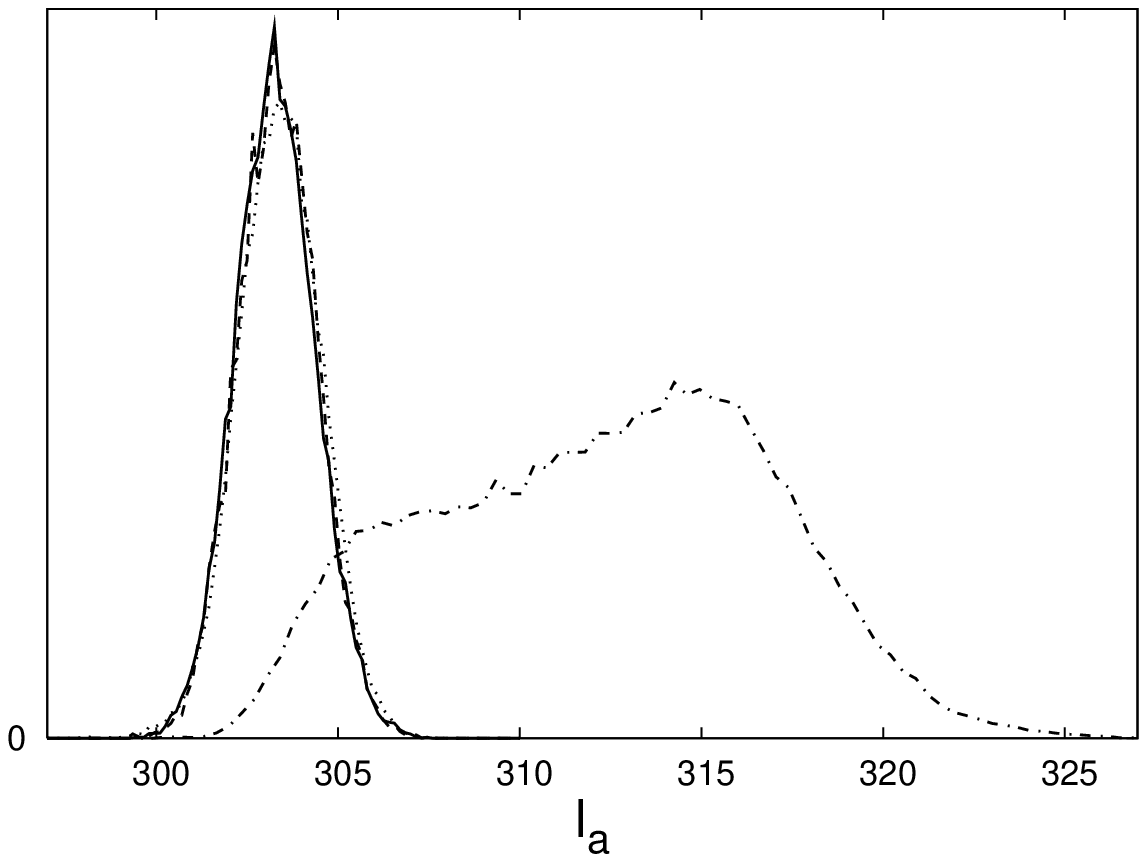}
\caption{The distribution of the shift parameter ${\cal R}$ and the 
acoustic scale $\ell _{\rm a}$ derived from MCMC chains with 
the WMAP data for four different models: $\Lambda$CDM (full line), dark energy with constant equation of state and dark energy perturbations (dotted line), open CDM (dashed line), and $\Lambda$CDM with massive neutrinos (dot-dashed line).}\label{fig2}
\end{center}
\end{figure}

In figure \ref{fig2} we show the distributions for 
${\cal R}$ and $\ell _{\rm a}$ when we assume a power-law primordial power 
spectrum, but make different assumptions about the matter and energy 
content of the Universe. Here the distributions are less scattered, and again we note
that the acoustic scale exhibits less variation than the shift parameter.
As a demonstration of using the shift parameter blindly, we also 
plot the distribution of ${\cal R}$ and $\ell _{\rm a}$
when we allow for massive neutrinos. This is clearly false since the size of the sound
horizon is now changed and hence the basis of using the shift parameter is no longer valid.
This is important to take note of, because we know that neutrinos do have a mass that 
should always be included as a free parameter in cosmological parameter estimation, 
even though it is commonly neglected.  

\section{Results}
Using the values of ${\cal R}$ and $\ell _{\rm a}$ 
calculated from the MCMC chains 
for the power-law $w$CDM model, we
can compare the resulting confidence contours with those arising 
from doing the full CMB fit. 
\subsection{$w$CDM model}
\begin{figure}[ht]
\begin{center}
\includegraphics[width=42mm,angle=0]{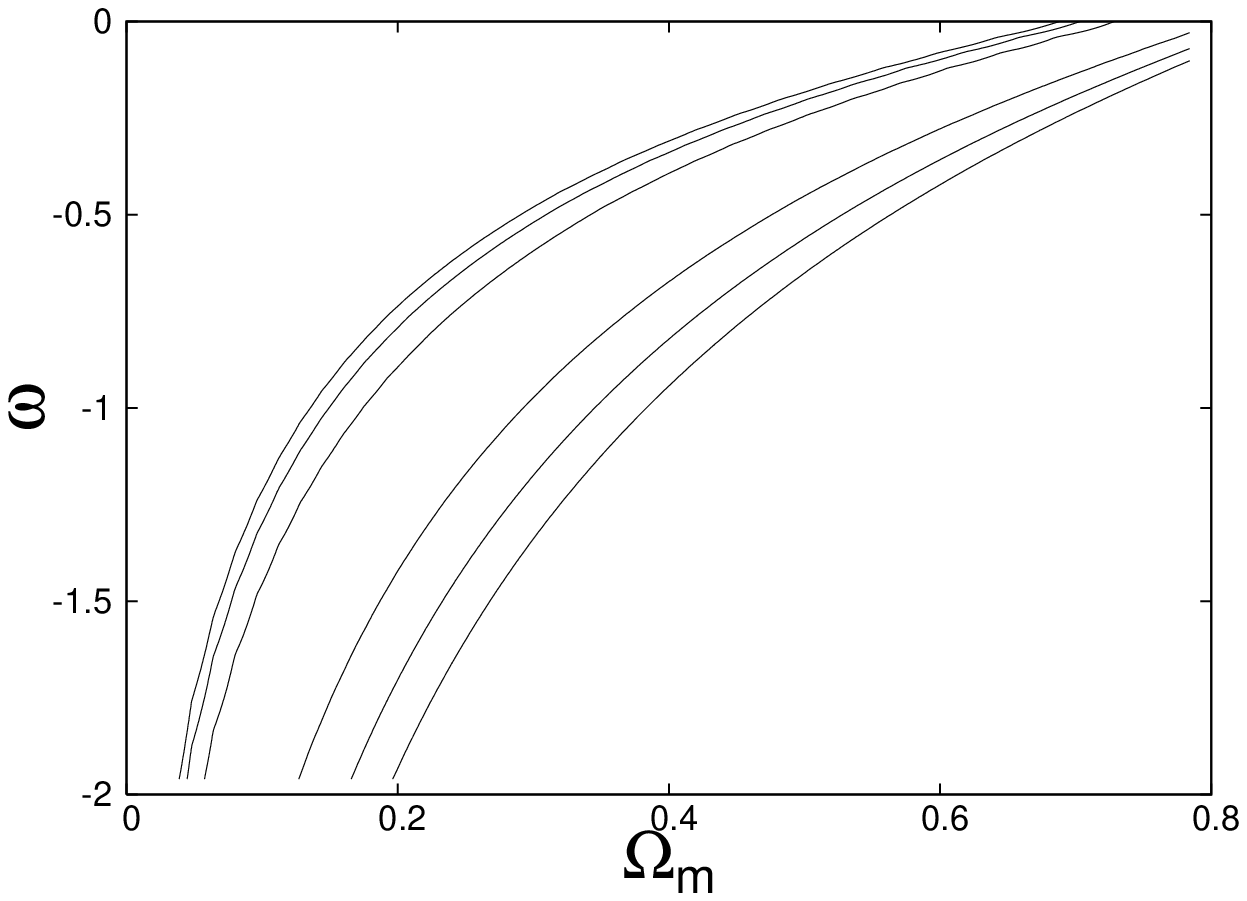}
\includegraphics[width=42mm,angle=0]{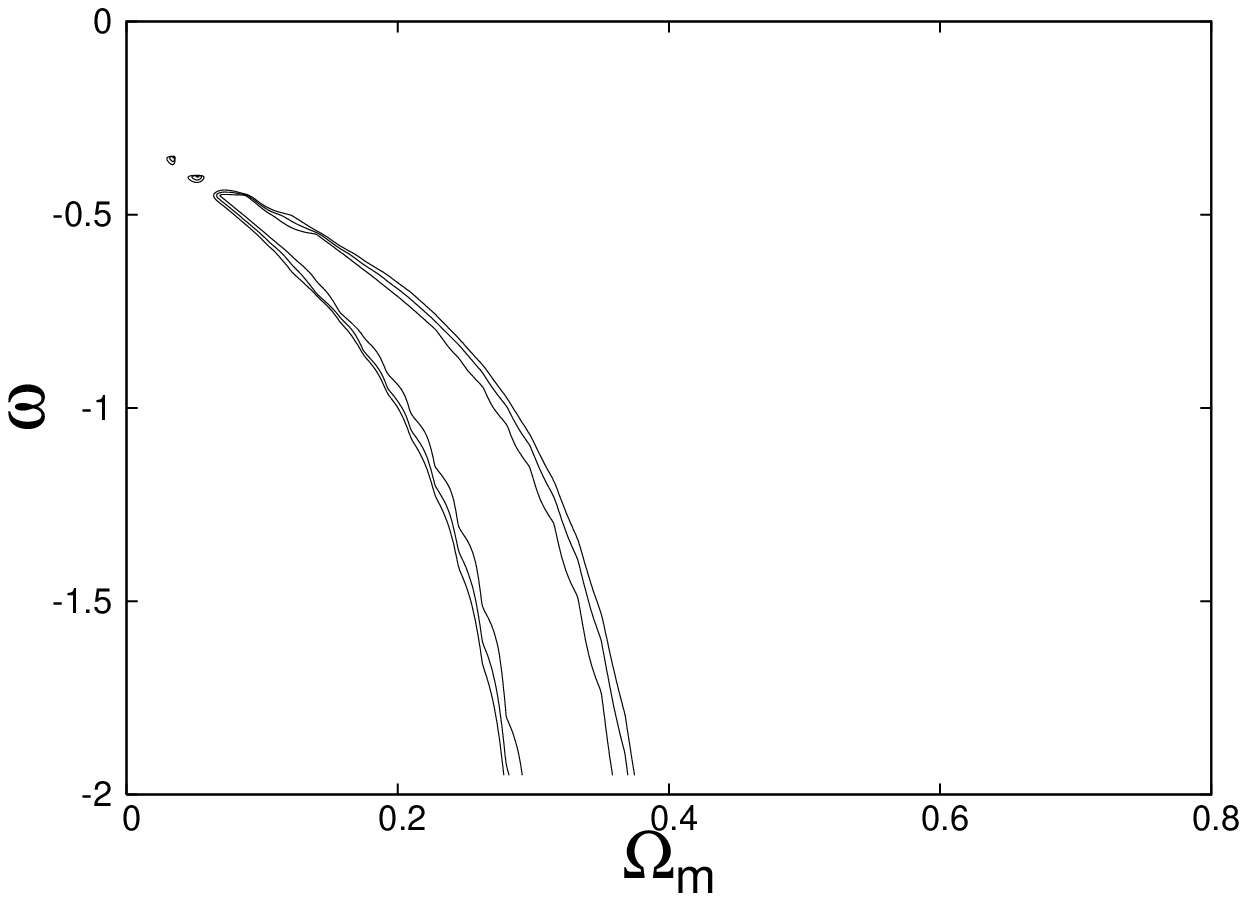}
\caption{The $68\%,\ 95\%$ and $99\%$ confidence levels calculated from
the shift parameter and acoustic peak position}\label{theory}
\end{center}
\end{figure}

\begin{figure}[ht]
\begin{center}
\includegraphics[width=50mm,angle=0]{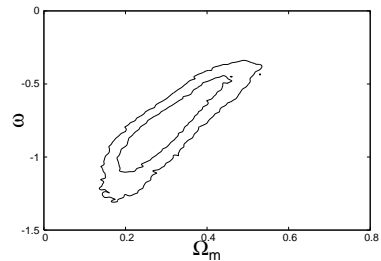}
\caption{Normalized probability density function from the WMAP
3-year data}\label{wmapnopert}
\end{center}
\end{figure}

In figure \ref{theory} we show the
$68\%,\ 95\%$ and $99\%$ confidence levels arising from using
the shift parameter and the acoustic peak position for the flat $w$CDM model.
In calculating the confidence limits using the acoustic peak position,
we have chosen a flat prior $h=0.73\pm0.03$ and marginalized over $h$.
We have kept the baryon density fixed at the WMAP value $\omega_{\rm b}=0.0223$
which we use throughout the paper unless otherwise stated.
Comparing this with the probability density plot from the full
MCMC chains, e.g., for the $w$CDM model with no dark energy perturbation, figure 
\ref{wmapnopert}, we see that the shift parameter gives a good approximation
to the CMB data while the acoustic peak position does not. Note that
even though $\omega_{\rm b}$ is fixed, ${\cal R}$ and $\ell_{\rm a}$ contours have fundamentally
different shapes, demonstrating the importance of the sound
horizon size in calculating $\ell_{\rm a}$.

The fact that the shift parameter approximates the full CMB contours
so well for the $w$CDM model is not surprising since the value of 
the shift parameter has been calculated from a chain that assumes 
that the cosmology is of the $w$CDM type. In other words, we 
first assume a model and then calculate chains that best fit the data
from which we derive a quantity. Following the same prescription
one can in fact construct other quantities that also well approximate
the full CMB contours, but are not physically motivated.

\subsection{Role of the Hubble parameter}
\begin{figure}[ht]
\begin{center}
\includegraphics[width=42mm,angle=0]{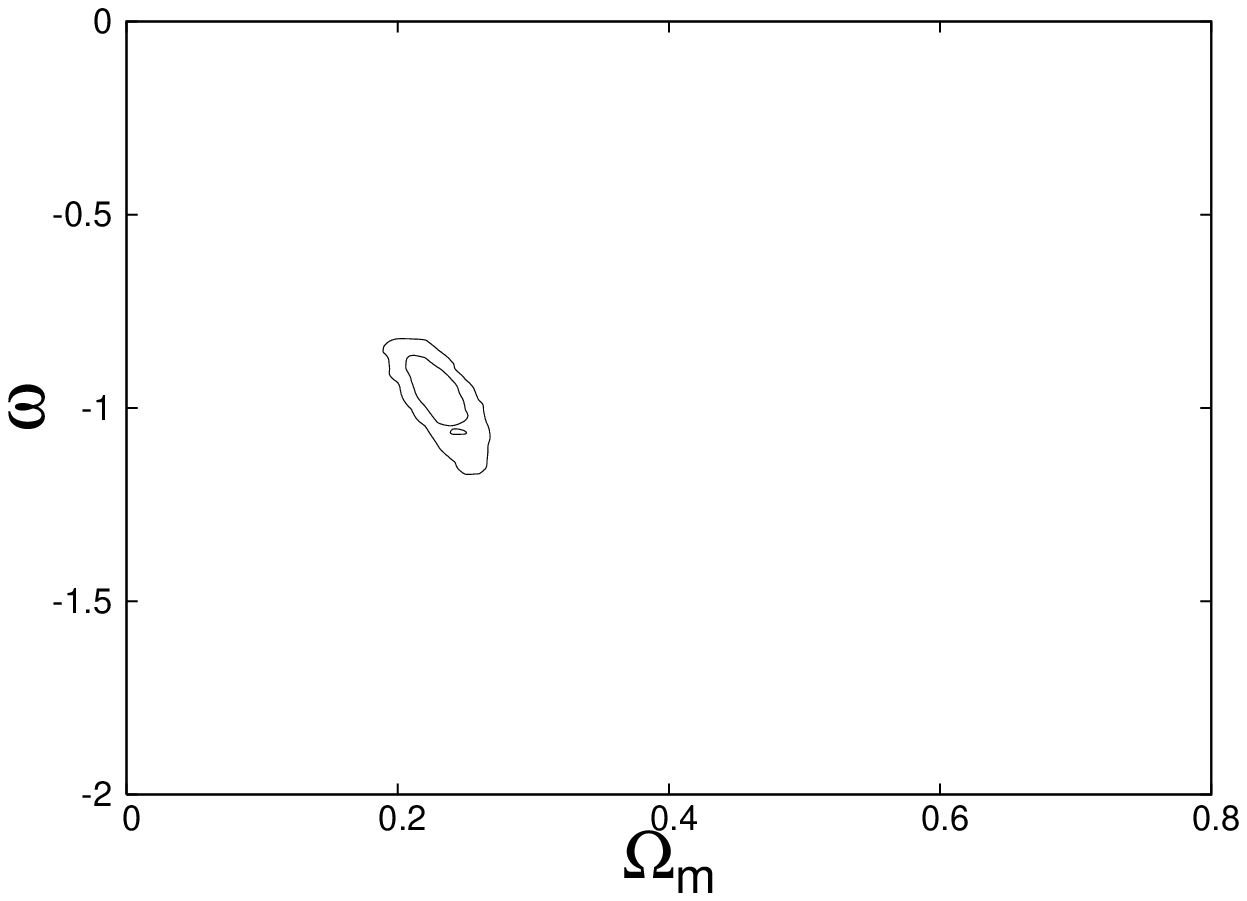}
\includegraphics[width=42mm,angle=0]{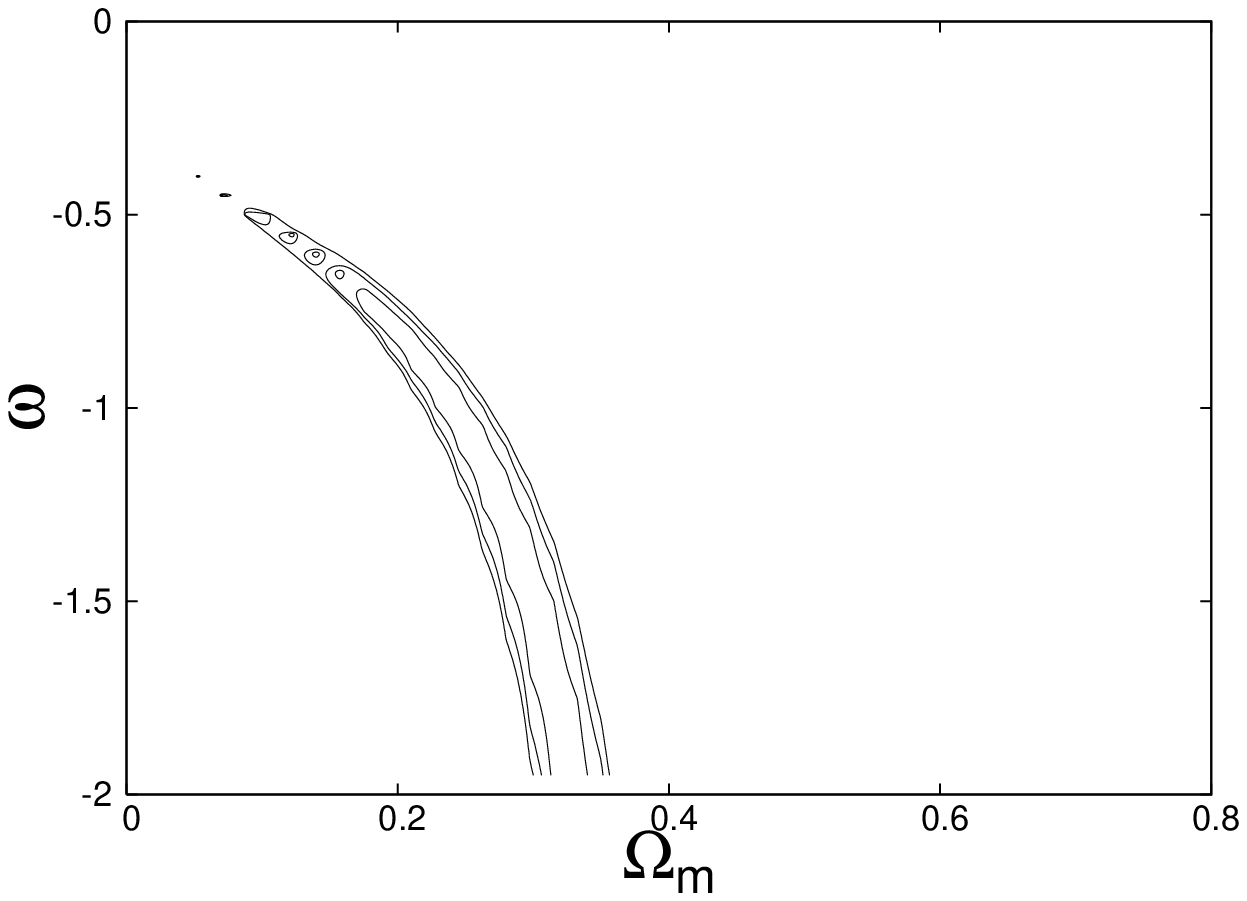}
\includegraphics[width=42mm,angle=0]{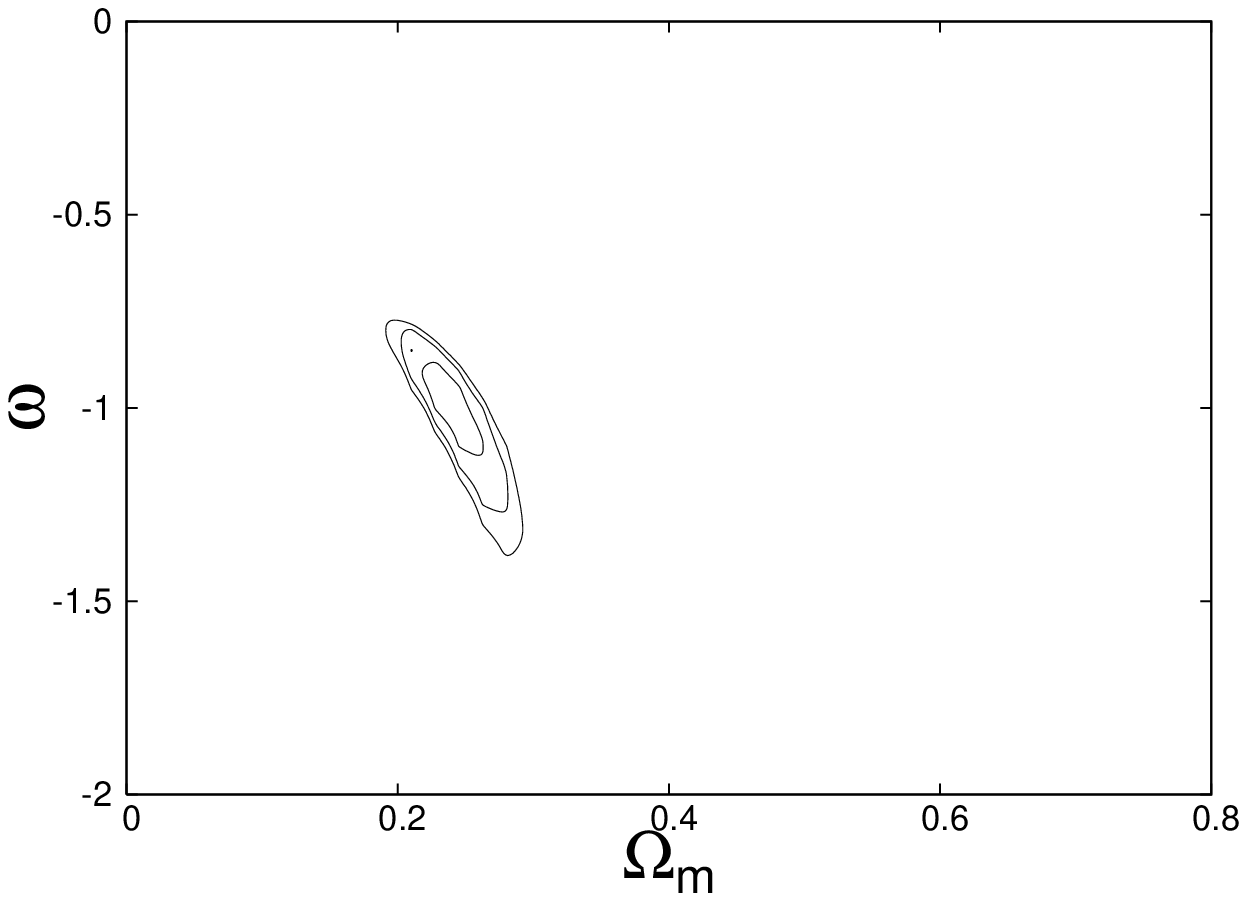}
\caption{Comparison of different methods: probability density from the wCDM MCMC chains (top left),
confidence contours calculated using $\ell_{\rm a}$ (top right) and the combined ${\mathcal R},\ \ell_{\rm a}$
contours (bottom).}\label{hfig}
\end{center}
\end{figure}

\begin{figure}[ht]
\begin{center}
\includegraphics[width=50mm,angle=0]{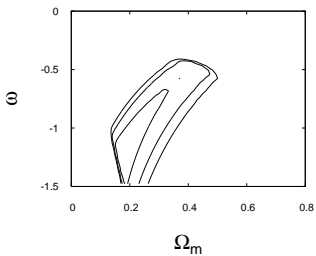}
\caption{The $68\%,\ 95\%$ and $99\%$ confidence levels calculated from
the combination of the shift parameter and $\ell_{\rm a}$, $h=0.5-1.0$}\label{combfig}
\end{center}
\end{figure}

In the WMAP chains, the Hubble parameter typically has a fairly large prior,
$0.5< h<1.0$. If the value of $h$ is constrained
by other observations, e.g., the Hubble Key Project reports $h=0.72\pm 0.08$
\cite{freedman}, the contours in the $(\Omega_{\rm m},w)$-plane look 
quite different. In figure \ref{hfig}, we show the normalized
probability density for the $w$CDM model from the WMAP provided chains 
(no dark energy perturbations) with cuts on $h$
along with the confidence contours calculated by using $\ell_{\rm a}$
with and without the shift parameter. The
shift parameter is independent of $h$  and hence the confidence contours are unchanged 
and shown in figure \ref{theory}.
From the figure one can conclude that 
the combination of ${\cal R}$ and $\ell_{\rm a}$ appears to be a good approximation
to the full CMB data when $h$ is constrained by independent observations.
This is further supported when we consider the confidence contours
arising from using the $({\cal R},\ell_{\rm a})$ combination with $h=0.5-1.0$ (flat prior), 
shown in figure \ref{combfig}.
Comparing to figure \ref{wmapnopert} we see that also in this case,
the combination of ${\cal R}$ and $\ell_{\rm a}$ is a reasonable approximation to the results 
obtainable from doing the full CMB fit, figure \ref{wmapnopert}.

\subsection{Role of baryons}
In the previous calculations, we have kept the baryon density fixed at the 
WMAP 3-year value, $\omega_{\rm b}=0.0223$.
This value is derived assuming the $\Lambda$CDM model and hence when using the 
${\cal R}$ or
$\ell_{\rm a}$ to study other cosmologies, one should be somewhat cautious when using this
value. A more robust, with respect to changing cosmology, measure of 
$\omega_{\rm b}$
comes from Big Bang nucleosynthesis (BBN), which gives 
$0.017\leq \omega_{\rm b} \leq 0.024$ ($95\%$ confidence level) with three massless neutrinos.
We find that changing the baryon density within the $95\%$ limits from BBN
has only a small effect on the results. In particular, when 
compared to the effect of changing the Hubble parameter, the significance of varying the
baryon density within the BBN limits is negligible.

\section{Non-standard cosmologies}

The shift parameter is particularly useful as a quick measure of how a
given cosmological model fits the CMB data.
In order to assess the validity of this approach, we compare here
the parameter constraints arising from the shift parameter and
from doing the full CMB fit on a non-standard model, namely on
a general Friedmann equation. Such a model is generalization of the
standard Friedmann equation and as such serves as a useful generic
no-standard model. We also consider the use of the acoustic peak position
as an useful approximation to the full CMB data fit.

\subsection{Modified Friedmann equation}
Modified Friedmann equations arise, e.g., in alternative theories of gravity. 
As an example, in the well known DGP model \cite{dvali}, 
the Friedmann equation on the brane is of the form 
\begin{equation}
H^2 \pm \frac{H}{r_{\rm c}} = \frac{8\pi G}{3}\rho_{\rm m},
\label{eq:modfried1}
\end{equation}
where $\rho_{\rm m}$ is the matter density on the brane and $r_{\rm c}$ is the cross-over scale
at which gravity starts to feel the effects of the fifth dimension.

Here we will consider modifications to the Friedmann equation with no spatial
curvature in the spirit of \cite{dvali2,elgaroy}.
A generalized Friedmann equation can be written as 
\begin{equation}
f(H,H_{\rm c})=H_0^2\Omega_{\rm m}(1+z)^3,
\label{genfried}
\end{equation}
where instead of modifying the matter content we consider 
modifications of gravity by having an arbitrary function $f$. 
The critical scale, $H_{\rm c}$, is close to the present Hubble parameter, 
$H_0$,
and determines when modifications from the standard
Friedmann equation start to have an effect. At early times, when $H\gg 
H_{\rm c}$, 
we know from BBN constraints that $f(H)\approx H^2$.
Keeping this mind and expanding in terms of $H_c/H$ gives
\begin{equation}
H^2\left[1+\sum_{n=1}^{\infty}c_n\left({H_c\over H}\right)^n\right]=H_0^2\Omega_{\rm m}(1+z)^3, 
\label{genfried2}
\end{equation}
from which it is clear that non-standard effects only start to have an effect at late times
when $H\sim H_{\rm c}$. 
Expanding the sum gives
\begin{equation}
H^2\left[1+c_1{H_c\over H}+c_2\left({H_c\over H}\right)^2+...\right]=H_0^2\Omega_{\rm m}(1+z)^3.
\label{genfried4}
\end{equation}
In this form, one can interpret the cosmological constant as a second order
correction to the Friedmann equation while the first order correction
corresponds to the DGP model. Generally, the $n$th order correction for a flat universe
is hence
\begin{equation}
\left({H\over H_0}\right)^2=\Omega_{\rm m}(1+z)^3+(1-\Omega_{\rm m})
\left({H_0\over H}\right)^{n-2}.
\label{genfried5}
\end{equation}
The leading correction to the Friedmann equation was previously studied using current
CMB, SNIa and LSS data by Elgar{\o}y and Multam\"{a}ki (2005). 

\subsection{Confidence contours}
\begin{figure}[ht]
\includegraphics[width=42mm,angle=0]{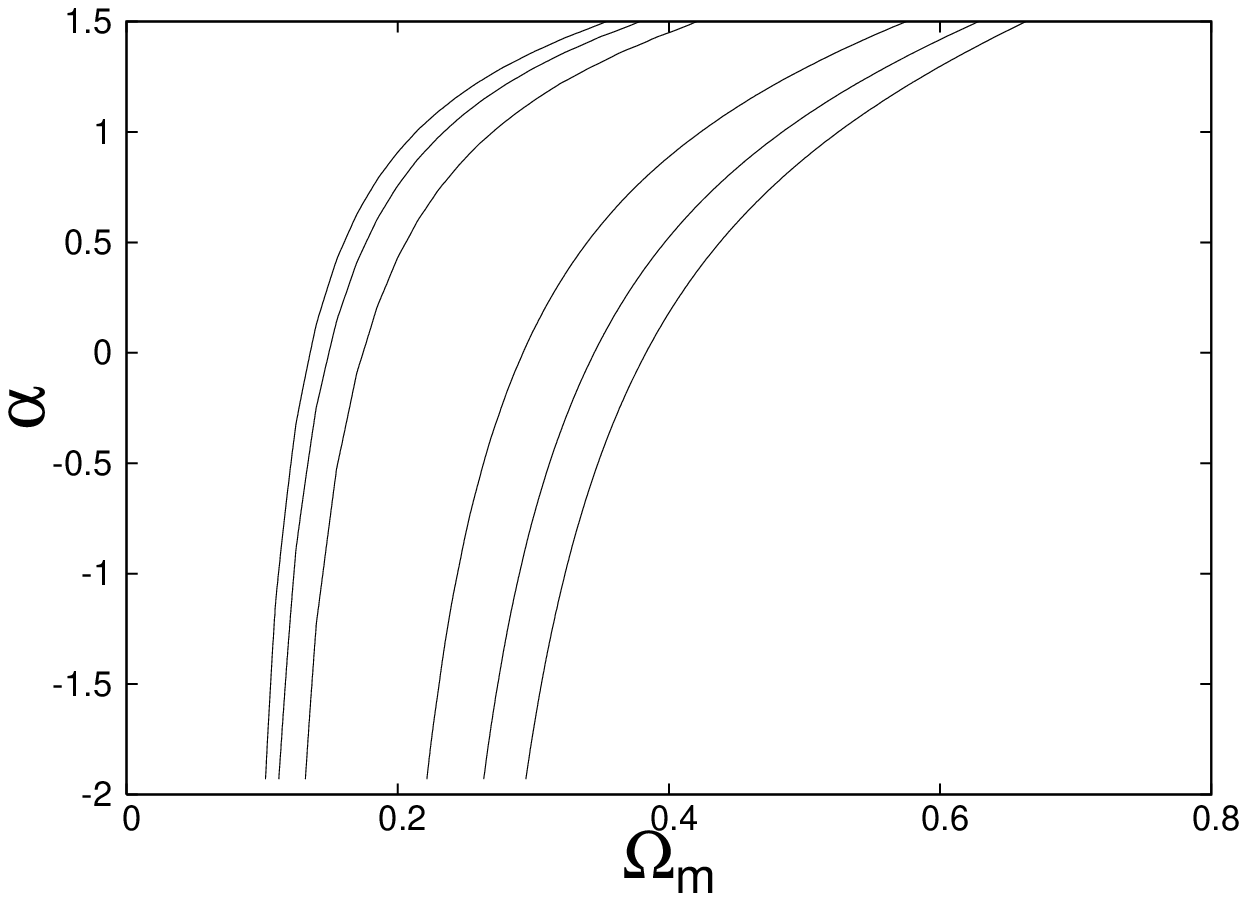}
\includegraphics[width=42mm,angle=0]{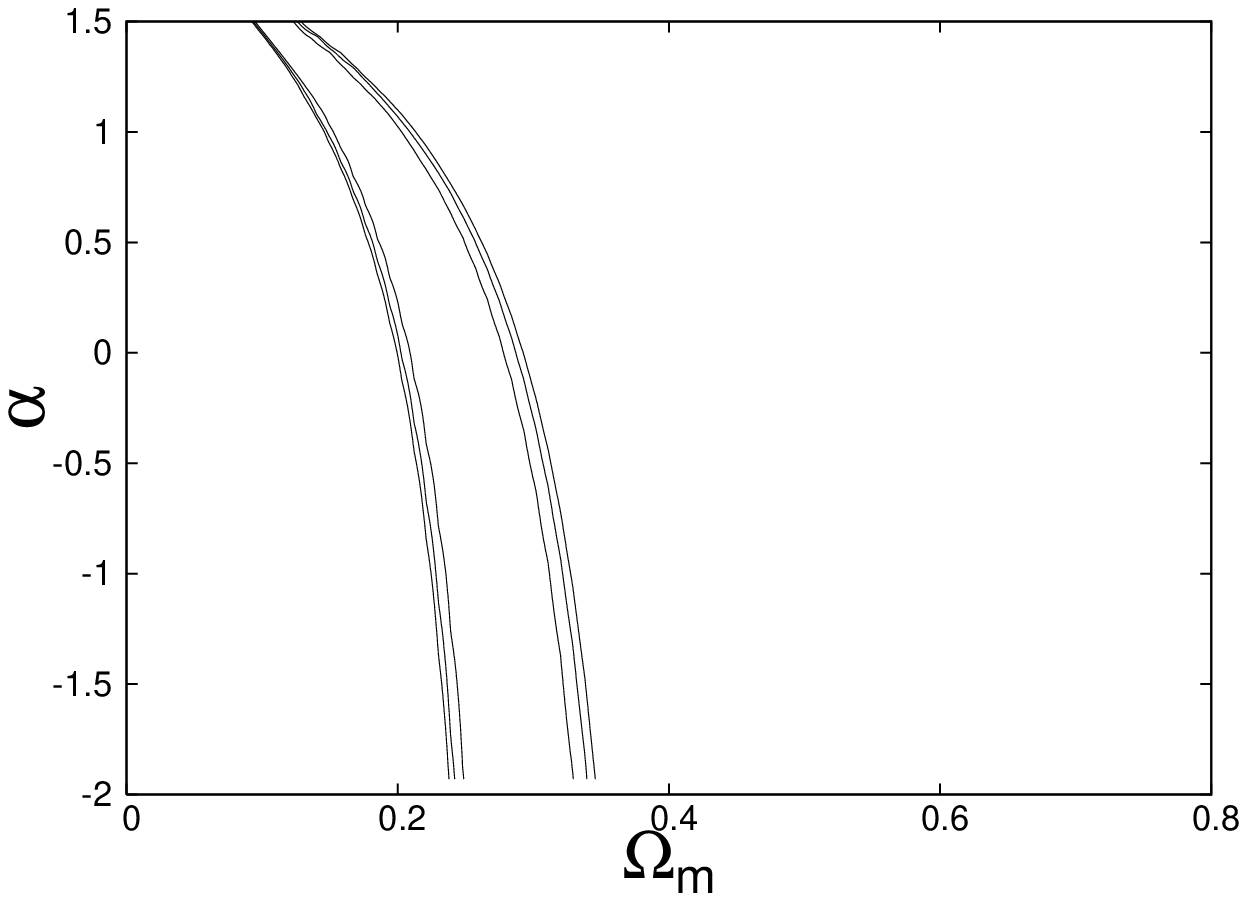}
\includegraphics[width=42mm,angle=0]{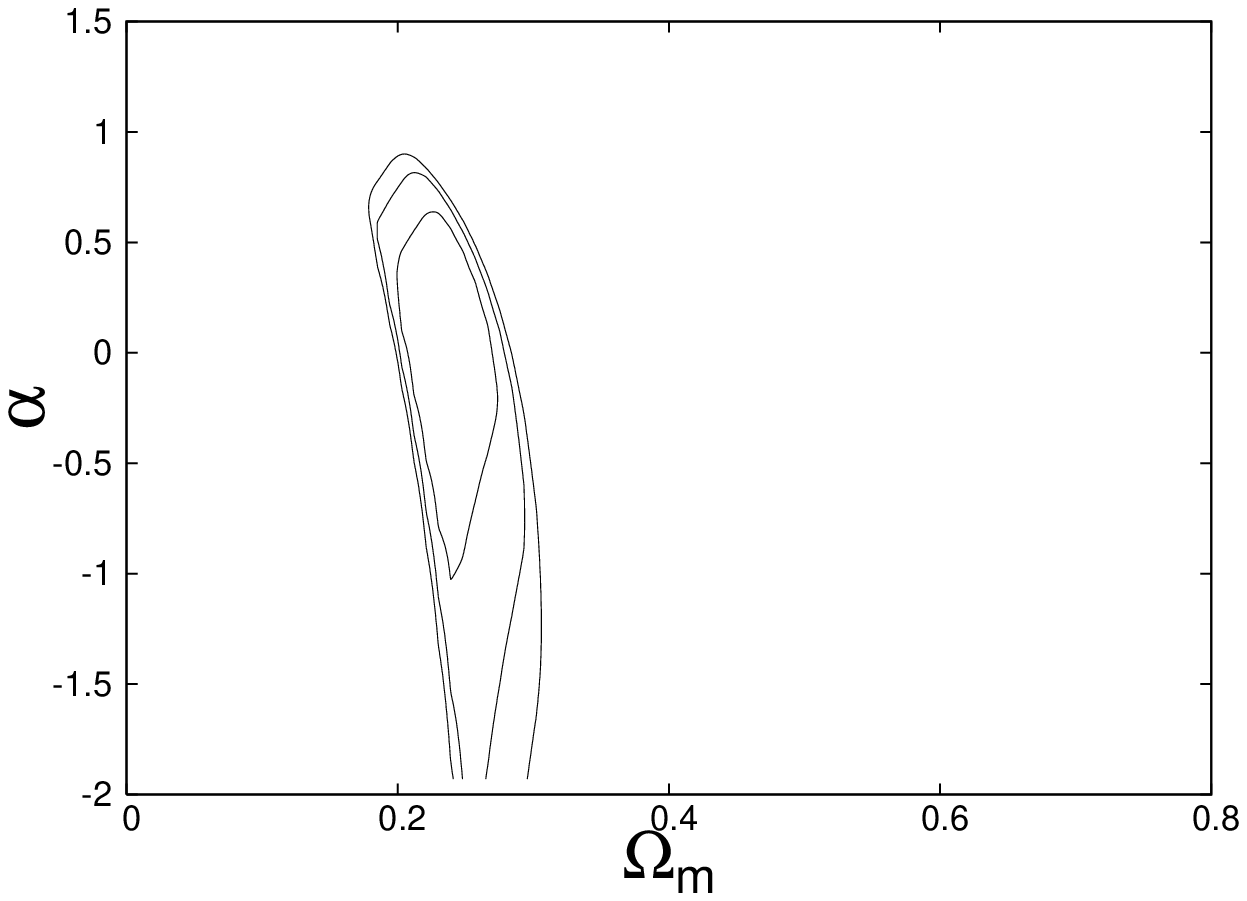}
\includegraphics[width=42mm,angle=0]{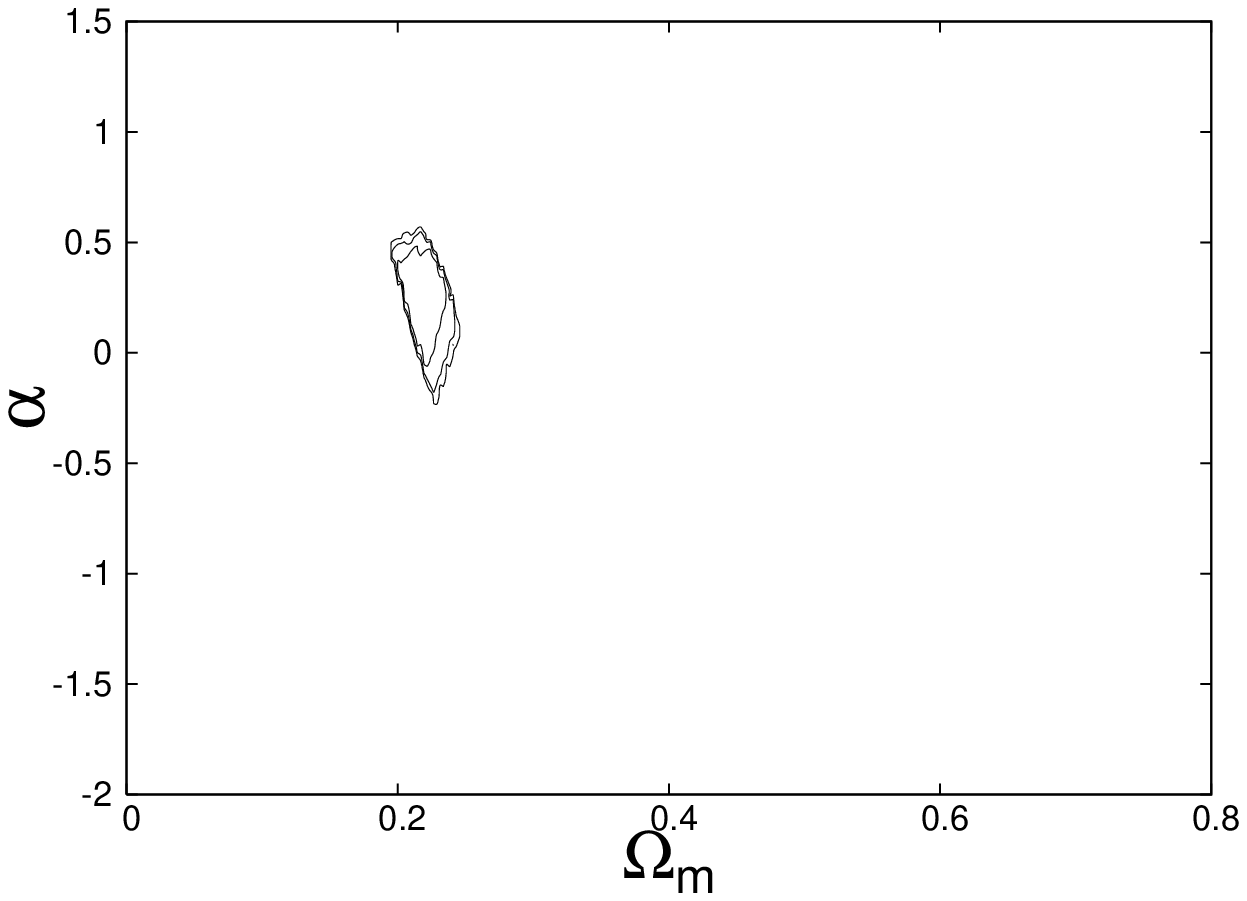}
\caption{The $68\%,\ 95\%$ and $99\%$ confidence levels calculated from the 
shift parameter (top left), acoustic peak position (top right), the combined ${\cal R}$ and 
$\ell_{\rm a}$ (bottom left) 
contours and from the full WMAP 3yr TT data (bottom right) ($h=0.73\pm0.03$, flat prior)}
\label{dtfig}
\end{figure}

The CMB spectrum arising within the context of the modified Friedmann equation, 
Eq. (\ref{genfried5}),
can be straightforwardly calculated by using, e.g., CMBFAST \cite{seljak} (see Elgar{\o}y \& Multam\"{a}ki (2005)  
for a detailed description) and fitted to the WMAP 3 year data. The resulting confidence
contours are shown in figure \ref{dtfig} along with contours from using the shift parameter
and the acoustic peak position. In figure \ref{dtfig} we also show the contours obtained
by combining the two approximate methods.
Again, we find that the combination of the acoustic peak along with the shift parameter gives
a good approximation to the full CMB fit.

\section{Conclusions}
In this work we have reconsidered the use of the shift parameter as a quick and 
easily implementable probe of dark energy. We find that while it gives an excellent
measure of the CMB spectrum for the $\Lambda$CDM model, 
with $\omega_{\rm b}$ and $\omega_{\rm c}$ fixed,
caution should be exercised when using the shift parameter to compare and constrain
non-standard cosmological models.

Even when using ${\cal R}$ to constrain more standard type models, such as the $w$CDM model,
careful consideration should be given to the value of the Hubble parameter, $h$. Since
the shift parameter is independent of $h$, but the full CMB spectrum fit is most 
definitely not, the shift parameter can be misleading when applied blindly.

The shift parameter is a geometrical measure as it measures the size
of apparent sound horizon at recombination. Keeping the sound horizon size fixed,
different cosmological models lead to different background expansion and hence
the shift parameter can be used to compare and constrain different models. However,
also the sound horizon size changes when varying cosmological parameters, most notably
changing the matter density, $\Omega_{\rm m}$ and the Hubble parameter, $h$. In addition,
massive neutrinos will also have an effect. Hence, in general the shift parameter will 
not be an accurate substitute for the CMB data and may in principle give misleading results 
when used to constrain non-standard results.

In addition, the value of the shift parameter used to constrain different dark energy models
is derived by first assuming the $\Lambda$CDM model, fitting the model to the data and then calculating
the value of ${\cal R}$. Again, for a general model, using the value obtained in this manner
is questionable since for a different model one may expect the shift parameter to be
different, while the CMB spectrum can fit well with observations.

In order to enhance the effectiveness of using the shift parameter as a cosmological tool,
we have considered adding information from the location of the first CMB peak, $\ell_{\rm a}$. 
Combining these two easily calculable observables, allows one to encompass information
from both the size of the sound horizon at recombination and the angular diameter distance
to it. As such, it more effectively constrains the allowed parameter space, including
the Hubble parameter that is not fixed when using only the shift parameter.
A possible caveat is, again, the fact that the numerical value of both of these parameters
is calculated within the $\Lambda$CDM framework, but by comparing to different models, we see that
the combination proves to be a good and efficient probe of non-standard cosmologies.

\begin{acknowledgements}
TM would like to thank the Institute of Theoretical Astrophysics, Oslo, for hospitality
during the completion of this work. This work has been supported by the Academy of Finland,
project no. 8111953.  {\O}E acknowledges support from the Research Council of 
Norway, project numbers 159637 and 162830. 
TM acknowledges the M-grid project, supported by the Academy of Finland, for computational resources.
We acknowledge use of the Legacy Archive for Microwave Background Data Analysis (LAMBDA).  Support 
for LAMBDA is provided by the NASA Office of Space Science. 
\end{acknowledgements}




\begin{thebibliography}{X}
\bibitem[Amarzguioui et al. 2006]{amarzguioui} Amarzguioui, M., Elgar{\o}y, 
{\O}., Mota, D. F. \& Multam\"{a}ki, T. 2006, A\&A, 454, 707
\bibitem[Astier et al. 2006]{astier} Astier, P. et al. 2006, A\&A, 447, 31
\bibitem[Barris et al. 2004]{barris} Barris, B. J. et al. 2004, ApJ, 602, 571
\bibitem[Bento et al. 2002]{bento} Bento, M. C., Bertolami, O. \& 
Sen, A. A. 2002, PhRvD, 66, 043507
\bibitem[Bilic et al.  2001]{bilic} Bilic, N., Tupper, G. B. 
\& Viollier, R. D. 2001, PhLB, 535, 17
\bibitem[Clocchiatti et al. 2006]{clocchiatti} Clocchiatti, A. et al. 2006, ApJ, 642, 1
\bibitem[Davis et al. 2007]{davis} Davis, T. M. et al. 2007, astro-ph/0701510
\bibitem[Dvali et al. 2000]{dvali} Dvali, G. R., Gabadadze, G. 
\& Porrati, M. 2000, PhLB, 485, 208
\bibitem[Dvali \& Turner 2003]{dvali2} Dvali, G. R. \& Turner, M. S. 2003, 
astro-ph/0301510
\bibitem[Efstathiou \& Bond 1999]{efstathiou} Efstathiou, G. \& Bond J. R. 
1999, MNRAS, 304, 75
\bibitem[Eisenstein et al. 2005]{eisenstein} Eisenstein, D. J. et al. 2005, 
ApJ, 633, 560
\bibitem[Elgar{\o}y \& Multam\"{a}ki 2005]{elgaroy} Elgar{\o}y, {\O}. 
\& Multam\"{a}ki, T. 2005, MNRAS, 356, 475
\bibitem[Freedman et al. 2001]{freedman} Freedman, W. L. et al. 2001, 
ApJ, 553, 47
\bibitem[Freese \& Lewis 2002]{freese} Freese, K. \& Lewis, M. 2002, 
PhLB, 540, 1
\bibitem[Gamerman \& Lopes 2006]{gamerman} Gamerman, D. \& Lopes, H. D. 
2006, Markov Chain Monte Carlo, Stochastic simulation for Bayesian inference 
(Boca Raton, Fla.: Taylor \& Francis)
\bibitem[Gondolo \& Freese 2003]{gondolo} Gondolo, P. \& Freese, K. 2003, 
PhRvD, 68, 063509
\bibitem[Hu \& Sugiyama 1996]{hu} Hu, W. \& Sugiyama, N. 1996, ApJ, 471, 542
\bibitem[Kamenshchik et al. 2001]{kamenshchik} 
Kamenshchik, A., Moschella, U. \& Pasquier, V. 2001, PhLB, 511, 265
\bibitem[Koyama 2006]{koyamaa} Koyama, K. 2006, JCAP, 0603, 017
\bibitem[Koyama \& Maartens 2006]{koyamab} Koyama, K. \& Maartens, R. 2006, 
JCAP, 0601, 016
\bibitem[Lazkoz et al. 2006]{lazkoz} Lazkoz, R., 
Maartens, R. \& Majoretto, E. 2006, PhRvD, 74, 083510
\bibitem[Lewis \& Bridle 2002]{lewis} Lewis, A. \& Bridle, S. 2002, PhRvD, 
66, 103511
\bibitem[Miknaitis et al. 2007]{miknaitis} Miknaitis, G. et al. 2007, 
astro-ph/0701043
\bibitem[Perlmutter et al. 1999]{perlmutter} Perlmutter, S. et al. 1999, 
ApJ, 517, 565
\bibitem[Riess et al. 1998]{riessa} Riess, A. G. et al. 1998, AJ, 116, 1009
\bibitem[Riess et al. 2004]{riessb} Riess, A. G. et al. 2004, ApJ, 607, 665
\bibitem[Sawicki et al. 2006]{sawicki} Sawicki, I., Song, Y.-S. 
\& Hu, W. 2006, astro-ph/0606285
\bibitem[Seljak \& Zaldarriaga 1996]{seljak} Seljak, U. \& Zaldarriaga, M. 
1996, ApJ, 469, 437
\bibitem[Song et al. 2006]{song} Song, Y.-S., Sawicki, I. \& 
Hu, W. 2006, astro-ph/0606286
\bibitem[Spergel et al. 2006]{spergel} Spergel, D. N. et al. 2006, 
astro-ph/0603449
\bibitem[Tonry et al. 2003]{tonry} Tonry, J. L. et al. 2003, ApJ, 594, 1
\bibitem[Wood-Vasey et al. 2007]{woodvasey} Wood-Vasey, W. M. et al. 
2007, astro-ph/0701041
\bibitem[Wang \& Mukherjee 2006]{wang} Wang, Y. \& Mukherjee, P. 2006, ApJ, 
650, 1
\bibitem[Wright 2007]{wright} Wright, E. L. 2007, astro-ph/0701584

\end{thebibliography}
\end{document}